\documentclass[
aps,%
11pt,%
final,%
notitlepage,%
oneside,%
twocolumn,%
nobibnotes,%
nofootinbib,%
superscriptaddress,%
noshowpacs,%
centertags]%
{revtex4}

\begin{document}

\title{A globular cluster in the dwarf galaxy Sextans~B}

\author{\firstname{M.~E.}~\surname{Sharina}}
\email{sme@sao.ru}
\affiliation{Special Astrophysical Observatory Russian Academy of Sciences,
N. Arkhyz, KChR, 369167, Russia}
\author{\firstname{T.~H.}~\surname{Puzia}}
\email{Thomas.Puzia@nrc.ca}
\affiliation{Herzberg Institute of Astrophysics, 5071 West Saanich Road,
Victoria, BC V9E 2E7, Canada}
\author{\firstname{A.~S.}~\surname{Krylatyh}}
\affiliation{Astronomy department of the Kazan State University,
Kremlevskaya street, Kazan, 420008, Russia}

\begin{abstract}
We present spectroscopic observations of a massive globular cluster in the
dwarf irregular galaxy Sextans~B, discovered by us on Hubble Space
Telescope Wide Field and Planetary Camera 2 (HST WFPC2) images. Long-slit
spectra were obtained with the SCORPIO spectrograph on the the 6-m
telescope at the Special Astrophysical observatory of the Russian Academy
of Sciences. We determine age, metallicity and alpha-element abundance
ratio for the globular cluster to be $ 2 \pm 1$ Gyr, $-1.35 \pm 0.25$ dex,
and $ 0.1 \pm 0.1$ dex, respectively. Main photometric and structural
parameters of it were determined using our surface photometry on the HST
images. The mass ($\sim 10^5 M \odot$), luminosity and structural
parameters appear to be typical for the globular clusters in our own
Galaxy. Our findings shed a new light on the evolutionary history of
Sextans~B.
\end{abstract}

\maketitle

\section{Introduction}
Discovery and detailed investigation of GCs in the smallest dwarf galaxies
is very important for development of a comprehensive theory of formation
and evolution of GCs. Elmegreen \& Efremov \cite{elmegrreen97} showed,
that massive gravitationally bound stellar clusters are possible
progenitors of the present-day old Galactic GCs, born in dense, compact
gas clouds in high ambient pressure environments. Such conditions were in
place in our Galaxy $\sim 10 \div 13$ Gyr ago, when the density and
pressure in interstellar and intergalactic medium were high. Observational
studies show, that the formation of the present-day massive
gravitationally bound stellar clusters can only happen near the centers of
massive galaxies, or in merging and interacting galaxies (e.g.
\cite{larsen00}).

Small, low surface brightness galaxies are often considered to be
survivors of ancient building blocks of galaxies according to the
generally accepted model of hierarchical galaxy formation. Unfortunately,
the formation of GCs in dwarf galaxies is not fully understood up to
now. The gas density is low in dwarf galaxies, and star bursts should lead
to enormous gas outflows due to shallow galactic gravitational potential
wells \cite{dekel}.

Studying of GCs in important for  a better understanding the
evolution of their host galaxies. There are three main morphological
types of dwarf galaxies. Irregular galaxies (dIr) are composed mainly of
young and intermediate-age stellar populations and are typically
found in the field and at the distance of $D > 300$ kpc from a nearby
massive galaxy \cite{grebel}. Spheroidal and elliptical dwarf galaxies
(dSph, dE) consist mainly of old stars \cite{grebel} and concentrate near
massive galaxies in groups and clusters \cite{einasto}.

Sextans~B is a rather isolated dwarf irregular galaxy, a member of the
Antlia-Sextans group near the Local group \cite{vdbergh99, tully02}. It is
a faint galaxy, typical for the population of the Local Group and other
nearby poor groups. Its central surface brightness in the filter B
of the wide-band Johnson-Cousins system is $22.8 \pm 0.2$ mag/sq.sec.
\cite{sh07}. The absolute blue magnitude, $M_B=-13.97$, corresponds to the
distance $1360 \pm 70$ kpc \cite{kara02}. The chemical composition of HII
regions and of planetary nebulae in the galaxy were extensively studied by
many authors \cite{stasinska86,skillman89,moles90,kniazev05}. The star
formation history of Sextans~B according to stellar photometry 
studies was rather complex. According to \cite{tosi91,sakai97,mateo98}
the galaxy experienced a powerful early star formation burst during the
first 1-2 Gyr, and an increased SF activity in the last 1-2 Gyr. Dolphin
et al. \cite{dolphin05} consider the star formation to be active during
the whole life of the galaxy. However, all the authors mention that the
available color-magnitude diagrams (CMDs) are not very deep to determine
the age of the galaxy with an accuracy of $\sim\!2-3$ Gyr. Kniazev
and co-authors \cite{kniazev05} discovered that one HII region in the
galaxy is twice as metal-rich in comparison to others. This implies
considerable inhomogeneity in the present-day metallicity distribution in
Sextans~B.

In this paper we report for the first time the discovery and a
detailed spectroscopic investigation of a GC in Sentans~B with the SCORPIO
spectrograph at the SAO RAS 6m telescope. Basic data for the GC obtained
in our work are summarised in Table~1:
(1),(2) - equatorial coordinates, (3) specific frequency of GCs in a
galaxy ($S_N = N \cdot 10^{0.4(M_V+15)}$, \cite{harris81}, where $N$ is a
number of GCs, $M_v$ is absolute V magnitude of the galaxy), (4)-(6) age,
metallicity and alpha element abundance ratio, (7) heliocentric radial
velocity, (8) integrated V magnitude, (9), (10) integrated absolute V
magnitude and color, corrected for Galactic extinction, (11) mass in solar
masses, (12)-(14) half-light radius, ellipticity, and projected distance
to the center of the galaxy; and model parameters obtained by fitting of
the surface brightness profile by the King law:
(15)-(18) central surface brightness in V and I, core and tidal radii.


\section{Globular cluster detection and photometry}
 A Hubble Space telescope Wide Field and Planet Camera 2 (HST WFPC2)
archive image of the GC in the V band (HST Proposal ID 8601) is
shown in the right panel of Figure~1. The cluster is partially
resolved into stars with a diffuse envelope. Observations were carried
out with the filters F606W and F814W, the band width of which are similar
to the filters V and I of the Johnson-Cousins system. The exposure 
time was 600 seconds in each filter.

The surface photometry and structural parameter determination for the GC
on the HST images were carried out using receipts described in detail in
our catalog paper of GCs in nearby dwarf galaxies \cite{catalog}.
The photometric results are shown in Figure~2 and summarised in Table~1.
The growth curves of integrated light in V and I bands (top) and the
distribution of the integrated color in magnitudes along the radius
(bottom) are shown in the left panel of the Fig.~2. Surface
brightness profiles in mag/sq.sec. in V and I (top) and the difference
between them (bottom) are shown on the right panel of Fig.~2. The
structural parameters obtained by fitting of the surface brightness
profile by the King law are similar to those of GCs in our own Galaxy.
However, the color of the GC in Sextans~B is blue, which indicates a
younger age. This statement will be  tested and quantified in
Section~4.


\section{Spectroscopic observations and data reductions}
The long-slit observations were conducted on February 11, 2005 with
the SCORPIO spectrograph \cite{afanasiev05}. The journal of the
observations is presented in Table~2. We used the grism VPHG1200g, the
spectral resolution 5 \AA\ , the spectral range 3800-5700 \AA, and
 the CCD-detector EEV42-40. The slit size was 6' x 1". The seeing was
about 2 arc seconds during our observations. We observed Lick/IDS and
radial velocity standard stars BF23751, HD115043 и HD132142 for proper
determination of radial velocities and calibration of our instrumental
absorption line measurements to the Lick standard system \cite{worthey94}.

The data reduction and analysis were performed using the European Southern
Observatory Munich Image Data Analysis System (MIDAS) \cite{banse83}. Cosmic-ray removal was done with the
FILTER/COSMIC program. We used the context LONG for the reduction of our
spectroscopic data. For each two-dimensional spectrum the standard
procedure of the primary reduction was done. After wavelength calibration
and sky subtraction, the spectra were corrected for extinction and
flux-calibrated using the observed the spectro-photometric standard
GRW+70$^{o}$5824 \cite{oke}. The dispersion solution provides the accuracy
of the wavelength calibration of $\sim$0.08 \AA. All one-dimensional
spectra of each object were summed to increase the S/N ratio.
The spectrum of the GC in Sextans~B is shown in Fig.3. The radial velocity
of the cluster was determined via cross-correlation with the radial
velocity standard stars. We used several radial velocity standards and
division on few regions of spectra for the determination of the radial
velocity error. The derived heliocentric radial velocity is listed in
Table~1. The signal-to-noise ratio per pixel in the spectrum of the
cluster measured at 5000 \AA\ is $\sim 40$. The quality of the spectrum
allows to accurately measure the absorption line indices and determine the
age, metallicity, and alpha-element abundance ratio ([$\alpha/Fe$] of 
the cluster by comparison of the measured values with predictions of
population synthesis model.

A detailed description of the Lick index measuring technique and of the
Lick/IDS system itself, as well as the demonstration of our tests on the
correspondence of our instrumental Lick index system to the standard one
are given in our methodological paper \cite{lickpaper}. In Table~3 the
averaged differences between the measured indices and the corresponding
indices in the Lick system, "c", are shown. We obtain zeropoints of
transformations of our index measurements into the Lick/IDS system by
averaging the values "c" for each index using all observed standard stars.
Note, that the calculated zeropoint transformations to the Lick/IDS system
agree well with those given in our methodological paper \cite{lickpaper}
for the multislit variant of observations with the same instrument.
Absorption-line indices for the GC in Sextans~B, measured by the GONZO
program \cite{puzia02} and transformed into the Lick system are listed in
the last column of Table~3.


\section{Age, metallicity and alpha element abundance ratio}
We determine age, $[Z/H]$ and $[\alpha/Fe]$ for the GC in Sextans~B using
our procedure of three-dimensional linear interpolation and $\chi^2$
minimization described in detail in \cite{M31dEGCs}.  This program
minimizes the difference between the measured Lick indices and the model
predictions \cite{thomas03} normalised to the index measurement
uncertainty. The program was extensively tested by comparison of ages,
$[Z/H]$ and $[\alpha/Fe]$ obtained by the routine with the same values
available in the literature for 12 Galactic and 46 M31 GCs
\cite{M31dEGCs}. The total Lick index measurement errors are determined
using the GONZO routine \cite{puzia02} from bootstrapping the object
spectrum. The errors include the Poisson errors and the radial velocity
measurement uncertainties. Note, that not only random index uncertainties,
but also errors of the transformations to the Lick system can alter the
measured ages, $[Z/H]$ and $[\alpha/Fe]$ ratios. Therefore, as many Lick
standard stars as possible should be observed to minimize the systematic
errors.

We determine age, metallicity, and $[\alpha/Fe]$ of the GC:
$2 \pm 1$ Gyr, $[Z/H]=-1.35 \pm 0.25$ dex, $[\alpha/Fe]=0.1 \pm 0.1$ dex.

A visual representation of how the Lick indices help to divide age and
metallicity effects can be obtained from the so-called age-metallicity
diagnostic plots \cite{puzia05} (see Figure~4). Lines in Figure~4
respresent the model predictions by Thomas et al. \cite{thomas03}. The
indices for the GC are shown by black dots. The rms errors of zeropoins of
the transformations into the Lick system are shown in the corners of the
panels. The plots show the dependence of age-sensitive H$\beta$,
H$\delta_A$, H$\gamma_A$, H$\delta_F$, H$\gamma_F$ on the [$\alpha/Fe$]
insensitive index [MgFe]'~$= \left\{ \mbox{Mg}b \cdot (0.72 \cdot
\mbox{Fe5270} + 0.28 \cdot \mbox{Fe5335})\right\}^{1/2}$. To explore the
Mg/Fe abundance ratio, the metal-abundance sensitive index $\langle
Fe\rangle = (Fe5335+Fe5270)/2$ is compared with the index $Mg_2$.


\section{Discussion}
It is interesting to compare the metallicities of different stellar
populations in Sextans~B. The metallicity of old stars, $[Fe/H]=-2.1$ dex,
was obtained by Grebel et al. \cite{grebel} via the comparison of the red
giant branch (RGB) with globular cluster fiducials. Star formation history
studies led to $[Fe/H] \sim -1.2 \div -1.3$ (\cite{tosi91, dolphin05}).
Our estimate of the GC's metallicity agree well with the last value. We
suggest that the time of the GC birth coincides with the strongest star
burst in the galaxy. However, the derived $[\alpha/Fe]=0.1$ likely implies
that there was a significant contribution from type-Ia SNe over the
past $\sim\!3$ Gyr. However, star formation was never so intense as in
giant elliptical galaxies, where $[\alpha/Fe]$ ratios for GCs can be
significantly higher up to $\sim\!0.5$ dex \cite{puzia06}. It is not
excluded, that the age of the parent galaxy is much higher than the age of
the GC. The colour-magnitude diagram of Sextans~B is deep enough to
clearly indicate the presence of red giants and a considerable number of
intermediate-age stars \cite{kara02}. However, the age of red giants is
not determined, and the presence of older stars was not proved. Dolphin et
al. \cite{dolphin03} pointed out, that the color of 13 Gyr old metal-poor
RGB stars is practically the same as the color of 2 Gyr old on
0.4$\div$0.8 dex more metal-rich stars.

We convert the $[Fe/H]$ value determined for stars and the GC in Sextans~B
to $[O/H]$ , assuming $[Fe/O] \sim 0$ \cite{skillman89} and
$12+log(O/H)_{\odot}=8.66$ \cite{asplund04}. The obtained value agrees
well with the one for the most HII regions in Sextans~B \cite{kniazev05}.
Thus, the metallicity of the GC and of young stellar populations in
Sextans~B appear to be similar. This fact does not contradict the
previous findings of star formation history studies which indicate a low
initial star formation activity. Additionally, Sextans~B could lose 
metals through stellar winds due to  a shallow gravitational
potential well.

 We can estimate the mass of the GC using the obtained age,
metallicity, luminosity, simple stellar population models of Bruzual and
Charlott \cite{bruzual03}, and  an assumed Salpeter initial mass
function. The obtained mass $0.8 ^{-0.25}_{+0.40} \cdot 10^5$, and
structural parameters, tidal and core radii (see table~1), are typical for
GC population of our Galaxy.

The specific frequency, $S_N \sim 3.8$, is much higher than the expected
one, if GCs form in direct proportion to the mass of the galaxy
\cite{laugh99}: $S_N \simeq \epsilon (1+M_{gas}/M_*) \simeq 0$, where
$M_{gas}$ is the mass of gas, $M_*$ is the mass of stars, and
$\epsilon=0.0025$ is a ratio of GC mass to the summary mass of gas and
stars. The relation $M_{gas}/M_* \sim 0.9 $ for Sextans~B can be estimated
using the hydrogen mass to luminosity ratio, $M_{HI}/L_B \sim 1.5$
\cite{springob05,kara04}, the stellar mass to to luminosity ratio,
$M_{*}/L_B \sim 1.58$ \cite{bell01}, and integrated color of the galaxy,
corrected for the Galactic reddening $(B-V)_0 \sim0.5$.
 The $S_N \sim 3.8$ is slightly lower than the expected one in the
case of mass loss with stellar winds \cite{dekel,harris01}.

Sextans~B is often compared in the literature with Sextans~A, the galaxy
similar  in structure, size, luminosity and the distance to
neighbours, which is also a member of the Antlia-Sextans group (e.g.
\cite{kniazev05}). It is interesting to note, that we did not find GCs in
the central part of this galaxy on the HST images.

It is unlikely, that the galaxy was influenced by ram pressure, tidal
stripping, and/or encounters with neighbors, because of its rather
isolated location. Hence, the star formation in Sextans~B was regulated
by internal mechanisms. Deep CMD studies and the detailed chemical
evolutionary modelling would  help to elucidate the nature of SF
activity and the GC formation mechanisms in the galaxy.


\section{Conclusions}
In this paper we report on the discovery and determination of
fundamental, evolutionary, photometric, and structural parameters of
a massive GC in the nearby dwarf irregular galaxy Sextans~B.

From the comparison of  population synthesis models \cite{thomas03}
 with the measured Lick absorption line indices we obtain an 
age, metallicity and [$\alpha/Fe$]: $2\pm 1$ Gyr, $[Z/H]=-1.35 \pm 0.25$
dex, and [$\alpha/Fe$]$=0.1 \pm 0.1$ dex. The metallicity of the GC agrees
well with the known in the literature for stars, HII regions and planetary
nebulae in the galaxy. The age of the GC coincides with the epoch of the
most powerful star formation activity in Sextans~B according to the
literature photometric studies. The alpha-element abundance ratio implies
a significant contribution of type-Ia SNe over the past $\sim\!3$ Gyr.

The  structural parameters and luminosity of the GC are typical
for Milky Way GCs. The mass of the GC is ($0.8 ^{-0.25}_{+0.40} \cdot 10^5
M \odot$), the core radius is $r_c = 1.7 \pm 0.15$ pc, and the tidal
radius is $r_t=40 \pm 2$ pc.

 The number of GCs in Sextans~B per unit luminosity is much higher
 than the expected one if  we assume that GCs form in the direct
proportion to the parent galaxy mass.

 The results of this work are important for solving the problem of
the origin of GCs in dwarf galaxies.  Our data indicate that massive
GCs can form in extremely shallow potential wells and isolated field
environments.


\begin{acknowledgments}
We thank Dr.~S.~N.~Dodonov for supervision of our observations, and Dr.~S.A.~Pustilnik for helpful discussions. T.H.P. gratefully acknowledges the
financial support through a Plaskett Research Fellowship from National
Research Council of Canada at the Herzberg Institute of Astrophysics.
\end{acknowledgments}

\pagebreak

\onecolumngrid

\newpage
\begin{figure*}[t!]
\setcaptionmargin{5mm}
\onelinecaptionstrue
\captionstyle{normal} \caption{DSS2-R 5x5 arcsec. image of Sextans~B
with marked the long slit position (left); WFPC2 HST image of the GC
(right).}
\end{figure*}
\begin{figure*}
\setcaptionmargin{5mm}
\onelinecaptionsfalse
\includegraphics[width=16cm]{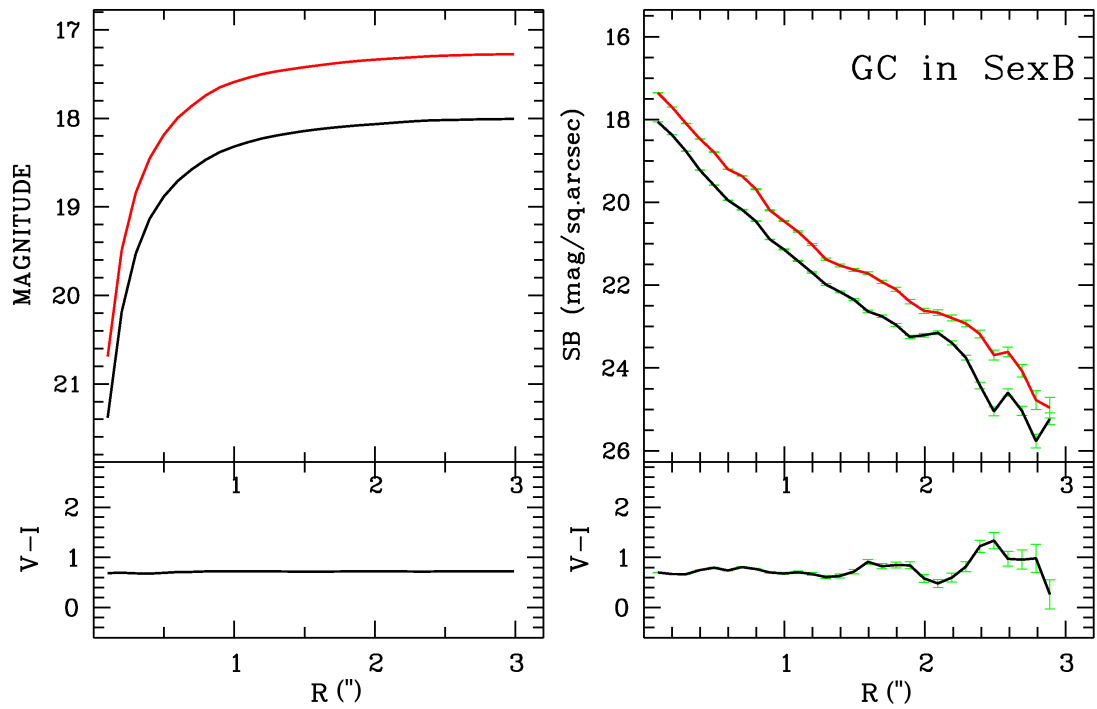}
\captionstyle{normal} \caption{Surface photometry results for
the GC in Sextans~B on the HST WFPC2 images.}
\end{figure*}
\newpage
\begin{figure*}
\setcaptionmargin{5mm}
\onelinecaptionsfalse
\includegraphics[width=10cm,angle=-90]{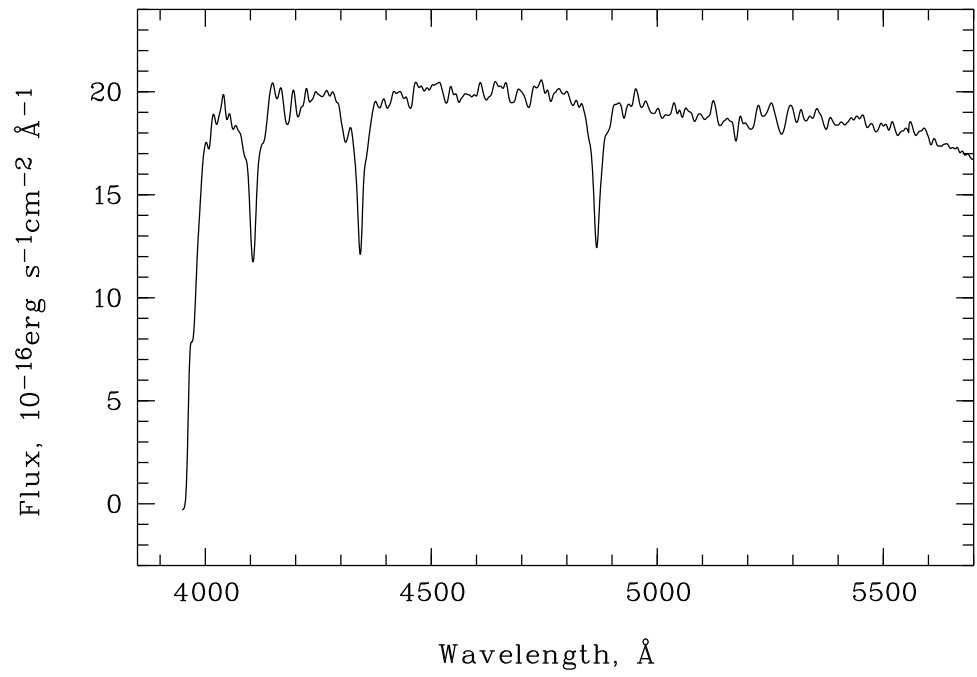}
\captionstyle{normal} \caption{The spectrum of the GC in Sextans~B
obtained with the SCORPIO spectrograph.}
\end{figure*}
\begin{figure*}
\setcaptionmargin{5mm}
\onelinecaptionsfalse
\includegraphics[width=16cm]{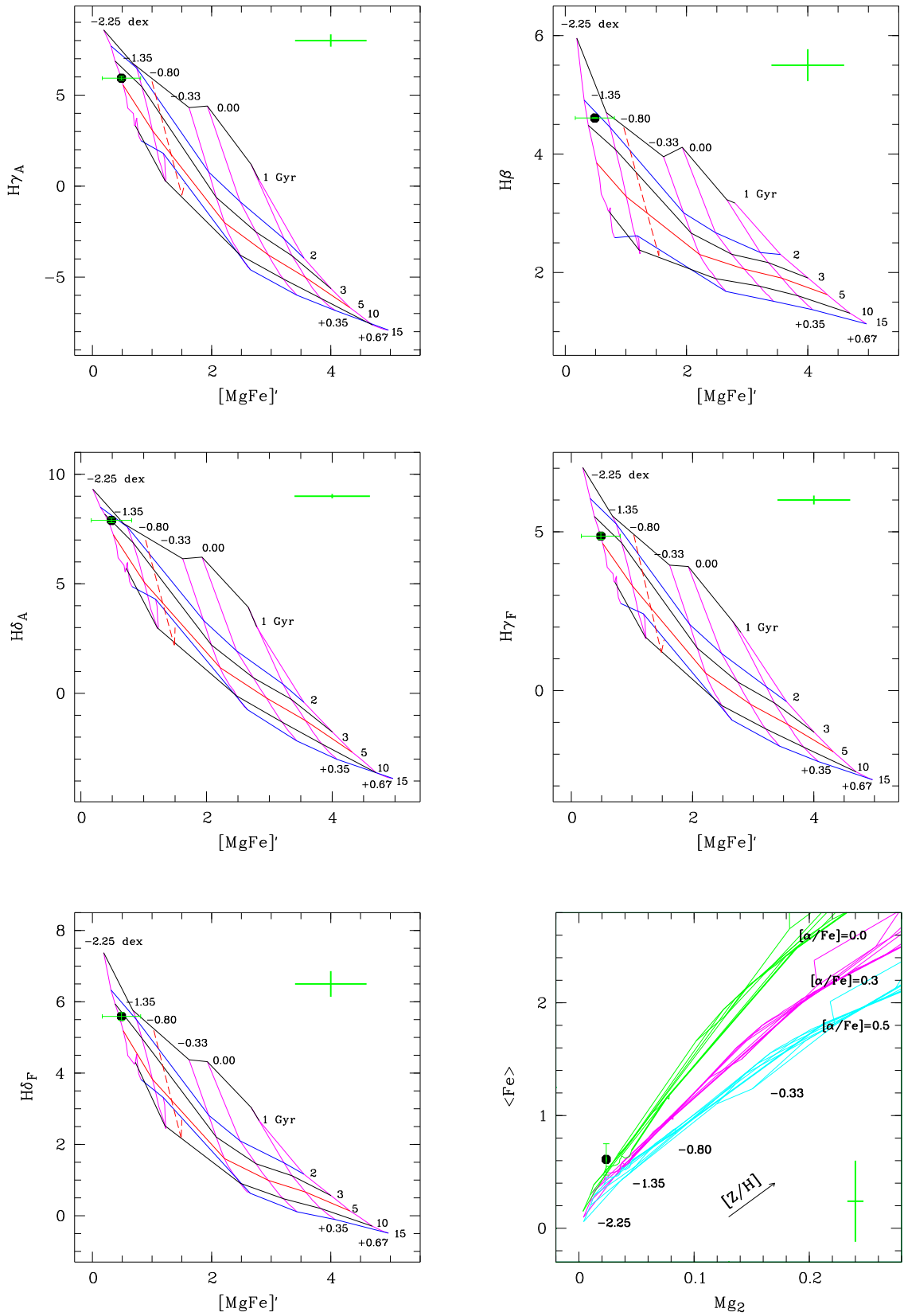}
\captionstyle{normal} \caption{Age-metallicity diagnostic plots for the GC in Sextans~B
The SSP model predictions from \cite{thomas03} are overplotted by lines.}
\end{figure*}

\pagebreak
\begin{table}[p]
\setcaptionmargin{0mm} \onelinecaptionsfalse
\captionstyle{flushleft} \caption{
Fundamental characteristics of the GC in Sextans~B determined in our work
(see the column content at the end of the section~1).}
\bigskip
\begin{tabular}{|r|l|c|}
\hline \hline
N & Parameter                  & GC     \\
\hline \hline
1 & RA(2000.0)                 & 10 00 04.64   \\
2 & DEC(2000.0)                & +05 20 07.4   \\
3 & $S_n$                      & 3.8            \\
4 & Age, Gyr                   & 2 $\pm$ 1        \\
5 & $[Z/H]$, dex               & -1.35 $\pm$ 0.3  \\
6 & $[\alpha/Fe]$, dex         & 0.1 $\pm$ 0.1    \\
7 & $V_h$, km/s                & 349$\pm$5      \\
8 & $V_{0}$, mag               & 17.90$\pm$0.02 \\
9 & $M_{V,0}$, mag             & -7.77$\pm$0.02 \\
10& $(V-I)_0$, mag             & 0.67$\pm$0.03  \\
11& Mass ($M \odot$)           & $0.8 ^{-0.25}_{+0.40} \cdot 10^5$  \\
12& $r_h$, pc                  & 4.1$\pm$0.2    \\
13& $e=1-b/a$                  & 0.05            \\
14& $d_{proj}$, kpc            & 0.45          \\
 \hline
15& $\mu_{V,0}$ mag/sq s       & 17.89$\pm$0.01  \\
16& $\mu_{I,0}$ mag/sq s       & 17.24$\pm$0.02  \\
17& $r_c$, pc                  & 1.7$\pm$0.15    \\
18& $r_t$, pc                  & 40$\pm$2        \\
\hline \hline
\end{tabular}
\end{table}

\begin{table}[p]
\setcaptionmargin{0mm} \onelinecaptionsfalse
\captionstyle{flushleft} \caption{Journal of spectroscopic observations}
\bigskip
\begin{tabular}{|l|c|c|}
\hline
Object     & Data       & Exposition   \\ \hline
GC in Sex~B& 11.03.2005 & 4x1200 s.     \\
GRW+70d5824& 11.03.2005 & 3x60          \\
HD115043   & 11.03.2005 & 10,20          \\
HD132142   & 11.03.2005 & 2x20           \\
HD2665     & 18.01.2007 & 21              \\
BF23751    & 11.03.2005 & 120            \\
\hline
\end{tabular}
\end{table}

\begin{table}[p]
\setcaptionmargin{0mm} \onelinecaptionsfalse
\captionstyle{flushleft} \caption{Correction terms of the transformation
to the Lick/IDS standard system \cite{worthey94}: $ I_{\rm Lick}=I_{\rm measured}+c $.
In the last column the GC indices corrected for the zeropoints of
transformations into the standard system are presented.}
\bigskip
\begin{tabular}{|l|c|c|c|c|}
\hline
Index      &   c    & rms error & units& Indices for GC in Sex~B \\ \hline
\hline
CN1        & -0.020 &  0.002    & mag  & -0.192 $\pm$0.001        \\
CN2        & -0.019 &  0.006    & mag  & -0.165 $\pm$0.001        \\
Ca4227     &  0.033 &  0.121    & \AA  & 0.410 $\pm$0.052        \\
G4300      &  0.786 &  0.720    & \AA  & 0.084 $\pm$0.057        \\
Fe4384     &  0.598 &  0.300    & \AA  & 0.040 $\pm$0.066         \\
Ca4455     &  0.288 &  0.030    & \AA  & 0.046 $\pm$0.068         \\
Fe4531     & -0.043 &  0.020    & \AA  & 1.718 $\pm$0.075         \\
Fe4668     & -0.163 &  0.210    & \AA  & 0.107 $\pm$0.085         \\
H$\beta$   & -0.909 &  0.270    & \AA  & 4.609 $\pm$0.085         \\
Fe5015     & 0.498  &  0.169    & \AA  & 2.692 $\pm$0.091         \\
Mg$_1$     & -0.035 &  0.013    & mag  & 0.020 $\pm$0.002         \\
Mg$_2$     & -0.019 &  0.007    & mag  & 0.024 $\pm$0.002         \\
Mgb        &  0.202 &  0.348    & \AA  & 0.336 $\pm$0.096         \\
Fe5270     &  0.554 &  0.358    & \AA  & 0.909 $\pm$0.098         \\
Fe5335     &  0.066 &  0.370    & \AA  & 0.312 $\pm$0.099         \\
Fe5406     &  0.001 &  0.010    & \AA  & 0.547 $\pm$0.100         \\
H$\delta_A$&  0.152 &  0.060    & \AA  & 7.897 $\pm$0.103         \\
H$\gamma_A$& -1.508 &  0.340    & \AA  & 5.931 $\pm$0.106         \\
H$\delta_F$& -0.345 &  0.360    & \AA  & 5.590 $\pm$0.107         \\
H$\gamma_F$& -0.069 &  0.142    & \AA  & 4.864 $\pm$0.108         \\
\hline
\end{tabular}
\end{table}

\begin{thebibliography}{99}
\bibitem{elmegrreen97}
B.G Elmegreen, Yu.N. Efremov, ApJ \textbf{480}, 235 (1997).
\bibitem{larsen00}
S. Larsen, T. Richtler, A\&A \textbf{354}, 836 (2000).
\bibitem{dekel}
A. Dekel, J. Silk, ApJ \textbf{303}, 39 (1986).
\bibitem{grebel}
E.K. Grebel, Gallagher III J.S., D. Harbeck, AJ \textbf{125}, 1926 (2003).
\bibitem{einasto}
J.Einasto, E. Saar, A. Kaasik, A.D. Chernin, Nature \textbf{252}, 111, (1974).
\bibitem{vdbergh99}
S. van den Bergh, ApJ \textbf{517}, L97 (1999).
\bibitem{tully02}
R.B. Tully, R.S. Sommerville, N. Trentham, M.A. Verheijen, ApJ \textbf{569}, 573 (2002).
\bibitem{sh07}
M.E. Sharina, Karachentsev I.D., Dolphin A.E. et al., MNRAS in prep. (2007)
\bibitem{kara02}
I.D. Karachentsev, M.E. Sharina, D.I.Makarov, A.E. Dolphin et al.,
A\&A \textbf{389}, 812 (2002).
\bibitem{stasinska86}
G. Stasinska, G. Comte, L. Vigroux, A\&A \textbf{154}, 352 (1986).
\bibitem{skillman89}
E.D. Skillman, R.C. Kennicutt, P.W. Hodge, ApJ \textbf{347}, 875 (1989).
\bibitem{moles90}
M. Moles, A. Aparicio, J. Masegosa, A\&A \textbf{228}, 310 (1990).
\bibitem{kniazev05}
A.Y. Kniazev, E.K. Grebel, S.A. Pustilnik, A.G. Pramskij, D.B. Zucker,
AJ \textbf{130}, 1558 (2005).
\bibitem{tosi91}
M. Tosi, L. Gregio, G. Marconi, P. Focardi, AJ \textbf{102}, 951 (1991).
\bibitem{sakai97}
S. Sakai, B.F. Madore, W.L. Freedman, ApJ \textbf{480}, 589 (1997).
\bibitem{mateo98}
M. Mateo, ARA\&A \textbf{36}, 435 (1998).
\bibitem{dolphin05}
A.E.Dolphin, D.R. Weisz, E.D. Skillman, J.A. Holtzman, astro-ph/0506430.
\bibitem{harris81}
W.E. Harris, S. van den Bergh, AJ \textbf{86}, 1627 (1981)
\bibitem{catalog}
M.E. Sharina, T. H. Puzia, D.I. Makarov, A\&A \textbf{442}, 85 (2005).
\bibitem{afanasiev05}
V.L. Afanasiev, A.V. Moiseev, Astronomy Letters \textbf{31} 194 (2005).
\bibitem{worthey94}
G. Worthey, ApJS,  \textbf{95}, 107 (1994).
\bibitem{banse83}
K. Banse, Ph. Crane, Ch. Ounnas, D. Ponz // MIDAS, in Proc. of DECUS, Zurich, p. 87 (1983).
\bibitem{oke}
J.~Oke, AJ \textbf{99}, 1621 (1990).
\bibitem{lickpaper}
M. E. Sharina, V. L. Afanasiev, T. H. Puzia, AstL \textbf{32}, 185 (2006).
\bibitem{puzia02}
T.H. Puzia, R. P. Saglia, M. Kissler-Patig, et al., A\&A \textbf{395}, 45 (2002).
\bibitem{M31dEGCs}
M. E. Sharina, V. L. Afanasiev, T. H. Puzia, MNRAS \textbf{372}, 1259 (2006).
\bibitem{thomas03}
D. Thomas, C. Maraston, R. Bender, MNRAS \textbf{339}, 897 (2003).
\bibitem{puzia05}
T.H. Puzia, M. Kissler-Patig, D.Thomas, et al., A\&A  \textbf{439}, 997 (2005).
\bibitem{puzia06}
 T.H. Puzia, M. Kissler-Patig, P. Goudfrooij,  ApJ  \textbf{648}, 383 (2006),
\bibitem{dolphin03}
A.E. Dolphin, A.Saha, E.D. Skillman, R.C. Dohm-Palmer, E. Tolstoy, A.A. Cole, J.
S. Gallagher, J.G. Hoessel, M. Mateo,
AJ  \textbf{126}, 187 (2003).
\bibitem{asplund04}
M. Asplund, N. Grevesse, A. J. Sauval, C. Allende Prieto, D. Kiselman,
A\&A \textbf{417}, 751 (2004).
\bibitem{bruzual03}
G. Bruzual, S. Charlot, MNRAS  \textbf{344}, 1000 (2003).
\bibitem{laugh99}
D.E. McLaughlin, AJ  \textbf{117}, 2398 (1999).
\bibitem{springob05}
C.M. Springob, M.P. Haynes, R. Giovanelli, B.R. Kent, ApJS \textbf{160}, 149 (2005).
\bibitem{kara04}
I.D. Karachentsev, V.E. Karachentseva, W.H. Huchtmeier, D.I. Makarov, AJ \textbf{127}, 2031 (2004).
\bibitem{bell01}
E.F. Bell, R.S. de Jong, ApJ \textbf{550}, 212, (2001)
\bibitem{harris01}
{\it W.E. Harris} //
in Star Clusters, Saas-Fee Advanced Course 28. Lecture Notes 1998,
 Swiss Society for Astrophysics and Astronomy. Edited by L. Labhardt and B. Binggeli.
Published by Springer-Verlag, Berlin (2001)
\end{thebibliography}
\end{document}